\documentclass{webofc}
\usepackage[varg]{txfonts}   
\usepackage{hyperref}

\usepackage{xcolor}
\usepackage{color}

\newlength{\figwidth}
\newlength{\fighalfwidth}
\newcommand{\larsoft}{LArSoft\xspace}
\newcommand{\wc}{Wire-Cell\xspace}
\newcommand{\wct}{\wc\xspace{}Toolkit\xspace}
 
\definecolor{codegreen}{rgb}{0,0.6,0}
\definecolor{codegray}{rgb}{0.5,0.5,0.5}
\definecolor{codepurple}{rgb}{0.58,0,0.82}
\definecolor{backcolour}{rgb}{0.95,0.95,0.92}

\begin{document}
\title{Evaluation of Portable Acceleration Solutions for LArTPC Simulation Using Wire-Cell Toolkit}
%
%

\author{\firstname{Haiwang} \lastname{Yu}\inst{1} \fnsep\thanks{Corresponding Author. \email{hyu@bnl.gov}} \and
        \firstname{Zhihua} \lastname{Dong}\inst{2} \and
        \firstname{Kyle} \lastname{Knoepfel}\inst{3} \and
        \firstname{Meifeng} \lastname{Lin}\inst{2} \and 
        \firstname{Brett} \lastname{Viren}\inst{1} \and
        \firstname{Kwangmin} \lastname{Yu}\inst{2}
}

\institute{
Department of Physics, Brookhaven National Laboratory, Upton, NY 11973, USA
\and
Computational Science Initiative, Brookhaven National Laboratory, Upton, NY 11973, USA
\and
Scientific Computing Division, Fermi National Accelerator Laboratory, Batavia, IL 60510, USA
}

\abstract{
  The Liquid Argon Time Projection Chamber (LArTPC) technology plays an essential role in many current and future neutrino experiments. Accurate and fast simulation is critical to developing efficient analysis algorithms and precise physics model projections. The speed of simulation becomes more important as Deep Learning algorithms are getting more widely used in LArTPC analysis and their
 training requires a large simulated dataset. Heterogeneous computing is an efficient way to delegate computing-heavy tasks to specialized hardware. However, as the landscape of the compute accelerators is evolving fast, it becomes more and more difficult to manually adapt the code constantly to the latest hardware or software environments. A solution which is portable to multiple hardware architectures while not substantially compromising performance would be very beneficial, especially for long-term projects such as the LArTPC simulations. In search of a portable, scalable and maintainable software solution for LArTPC simulations, we have started to explore high-level portable programming
 frameworks that support several hardware backends. In this paper, we will present our experience porting the LArTPC simulation code in the Wire-Cell toolkit to NVIDIA GPUs, first with the CUDA programming model and then with a portable library called Kokkos. Preliminary performance results on NVIDIA V100 GPUs and multi-core CPUs will be presented, followed by a discussion of the factors affecting the performance and plans for future improvements.  
}
\maketitle

\section{Introduction}
\label{intro}

The Liquid Argon (LAr) Time Projection Chamber (LArTPC) is a key detector technology that is widely used in neutrino physics~\cite{Acciarri:2016smi, Abi:2017aow, Antonello:2015lea, Abi:2020evt}. Both scintillation light and ionization electrons are created when charged particles traverse the LAr medium.  The scintillation light can provide neutrino-interaction timing information while the detection of ionization electrons affords high-resolution position and energy information.  

Accurately simulating such detector responses is difficult due to the highly diverse event topologies that occur in LArTPCs.  The large number of disparate data patterns is (in part) responsible for the rising popularity in the neutrino community of deep-learning algorithms, which are capable of processing such data volumes. Training such algorithms, however, requires very large data sets to achieve accurate performance, presenting a challenge in terms of data handling and processing.  Improved computational performance for simulation is thus key to generating enough data in a timely manner.

This paper discusses some of the challenges associated with LArTPC simulation and some efforts undertaken to improve the efficiency of generating simulated detector signals.  We have made various optimizations, including using symmetries to simplify computing algorithms; trying different vendor libraries for the fast Fourier transform (FFT) calculations; and also using task-level parallelization with the Intel Threading Building Blocks (TBB)~\cite{tbb} to balance CPU and memory load.  To further boost efficiency, we are exploring using graphical processing units (GPUs) as accelerators with the NVIDIA CUDA~\cite{cuda} API and programming model.  Although the initial results are promising, two considerations motivate another approach:

\begin{enumerate}
    \item the hardware available to us may not have the specific type of accelerators (such as NVIDIA GPUs) the code is programmed for, and
    \item as technology evolves, new and better accelerators may become available.
\end{enumerate}
 
\noindent It is thus ideal to pursue a portable solution that can provide a unified user-level API while hiding the lower-level, accelerator-specific, backend interactions.  Following this strategy, we started our evaluation with Kokkos~\cite{kokkos}, which provides a set of unified APIs for multiple back-ends.  In this paper, we present our experience porting the \wc LArTPC simulation to Kokkos, including the impact on the framework, the algorithm and the kernel.

This paper is organized as follows. Section~\ref{sec-lartpc-sim} briefly summarizes LArTPC simulation and its implementation in \wct. The CUDA porting experience is described in Section~\ref{sec-cuda} and the Kokkos porting experience in Section~\ref{sec-kokkos} where we also present preliminary benchmark results. Future plans are described in Section~\ref{sec-kokkos-plan}. And we summarize in Section~\ref{sec-summary}. 
\section{LArTPC Simulation}
\label{sec-lartpc-sim}

\begin{figure}[thb]
  \centering
  \includegraphics[width=0.8\figwidth]{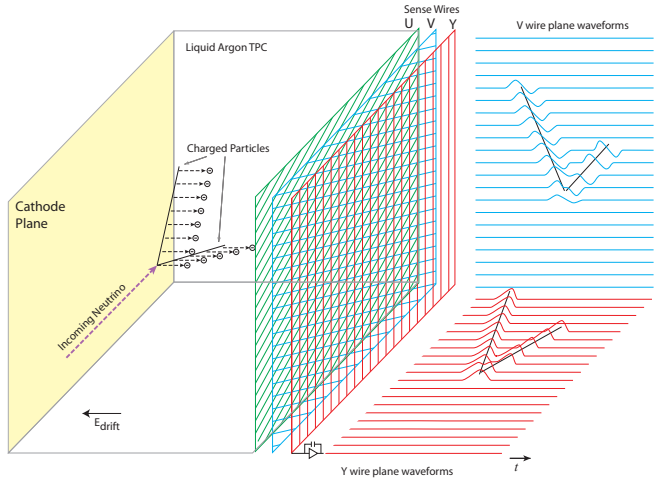}
  \caption{Diagram from ref.~\cite{Acciarri:2016smi} showing conceptual configuration of a typical three wire plane LArTPC and illustrating the signal formation of it. The signal in the U induction plane is omitted from the diagram for simplicity.}
  \label{fig:lartpc-concept}
\end{figure}

Figure~\ref{fig:lartpc-concept} illustrates the signal formation in a typical LArTPC configuration with wire read-out. When ionization electrons move close to the wires, induced currents can be detected. As governed by Ramo's theorem~\cite{ramo}, the induced current has bipolar and unipolar shapes on the induction-plane and collection-plane wires, respectively. The recorded digitized TPC signal can be modeled as a two-dimensional (2D) convolution of the distribution of the arriving ionization electrons and the impulse detector response: 
\begin{equation}\label{eq:conv}
  M(t,x) = \int_{-\infty}^{\infty}\int_{-\infty}^{\infty} R(t-t',x-x') \cdot S(t',x')dt' dx' + N(t,x),
\end{equation}
where $M(t,x)$ is a measurement, such as an analog-to-digital converter (ADC) value at a given sampling time, $t$, and wire position, $x$.
$R(t-t',x-x')$ is the impulse detector response, including both the field response that describes the induced current by a moving ionization electron and the electronics response from the shaping circuit. 
$S(t',x')$ is the charge distribution in time and space of the arriving ionization electrons, 
and $N(t,x)$ is the electronics noise.
The goal of TPC simulation is to calculate $M(t,x)$ based on the original charge distribution $S$ given the known detector response $R$ in the presence of the electronics noise $N$.
The integral in Eq.~\ref{eq:conv} is referred to as the \textit{signal simulation} and the additive term is referred to as the \textit{noise simulation}.  Computing the signal contribution is typically more time-consuming than computing the noise simulation.  References~\cite{Adams:2018dra,Adams:2018gbi} introduced a LArTPC detector response simulation algorithm based on the 2D convolution, which is considered the current state-of-the-art and widely used in multiple experiments.  

\subsection{\wct}
\subsubsection{Algorithm for signal simulation}
The \wct calculates the 2D convolution by transforming to the frequency domain, applying a multiplicative correction, and then transforming back to the time-space domain:
\begin{align}\label{eq:conv-freq}
\begin{split}
  S(t,x) \quad &\underrightarrow{\quad FT \quad} \quad S(\omega_{t},\omega_{x}), \\
  M(\omega_{t},\omega_{x}) &= R(\omega_{t},\omega_{x}) \cdot S(\omega_{t},\omega_{x}), \\
  M(\omega_{t},\omega_{x}) \quad &\underrightarrow{\quad IFT \quad} \quad M(t,x).
\end{split}
\end{align}
\noindent where $R(\omega_{t},\omega_{x})$ is the pre-calculated detector response function in the frequency domain.
The signal simulation process can be factored into two steps, $S(t,x)$ calculation and $M(t,x)$ calculation through Eq.~\ref{eq:conv-freq}.
The $S(t,x)$ calculation can be further divided into two sub-steps: 1) rasterize each energy deposition into small patches ($\sim 20 \times 20$), and 2) add up all the patches to a large grid ($\sim10k \times 10k$).
These two sub-steps will be referred as ``rasterization'' and ``scatter adding'' below.
The $M(t,x)$ calculation uses Fourier Transform as the main operation so will be referred as ``FT''.

\begin{figure}[thb]
  \centering
  \includegraphics[width=0.8\figwidth]{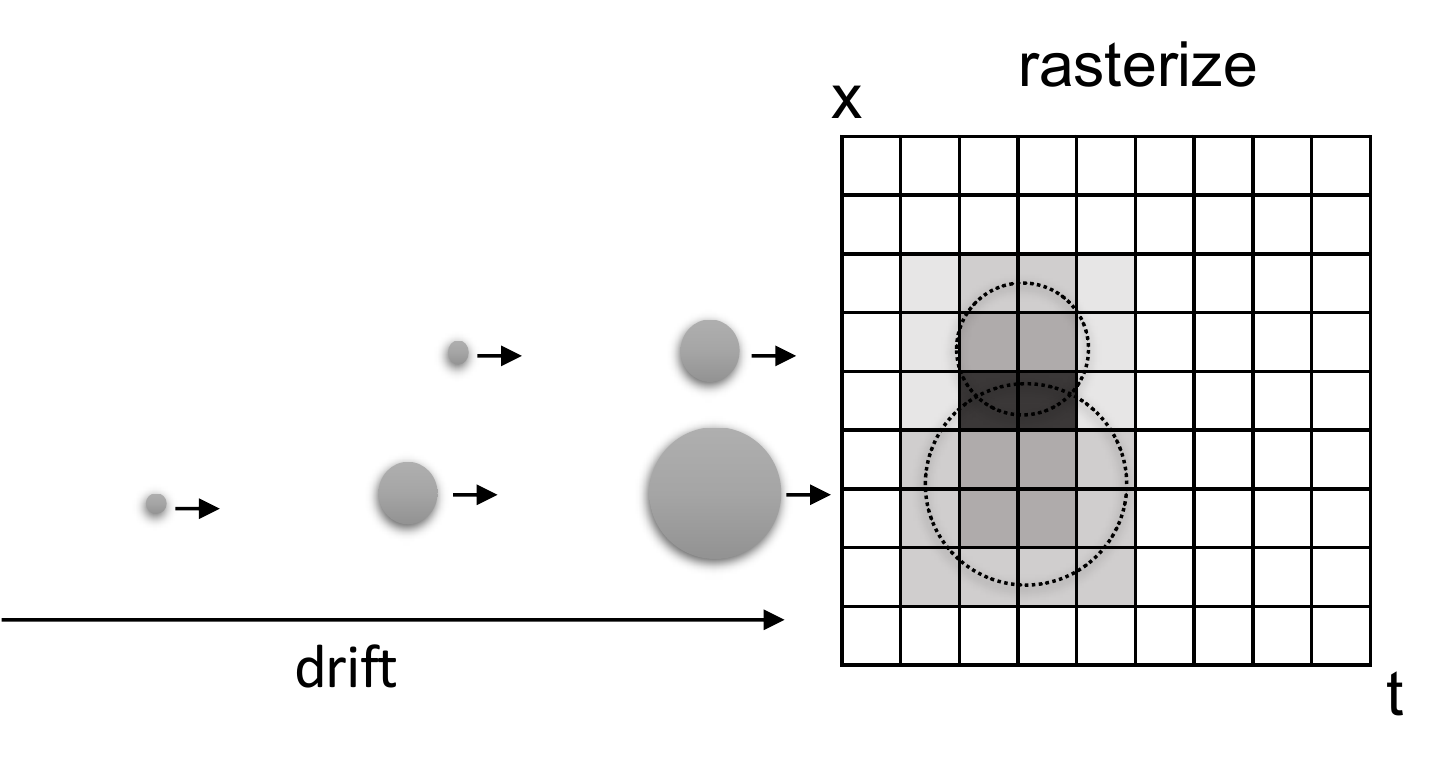}
  \caption{To illustrate the electron cloud drifting simulation and the rasterization process. In this figure, two electron clouds drift towards to the read-out plane. They expand in both transverse and longitudinal directions and overlap at the read-out plane.}
  \label{fig:rasterize}
\end{figure}

\subsubsection{Programming model and language}

Written primarily in C++, the \wct~\cite{wirecell_toolkit} is a software package designed according to the dataflow programming paradigm~\cite{10.1145/1013208.1013209}.  It supports a modular computing model by expressing computing tasks as nodes of a graph.  These nodes are connected to form directed acyclic graphs that can be executed by various processing engines.  The nodes themselves are polymorphic, allowing toolkit users to create and assemble concrete components according to a common framework interface.  In addition to the framework infrastructure, the toolkit also provides many concrete algorithms for specific LArTPC analysis steps.  The current state-of-the-art 2D convolution-based simulation is one of them, which can be run standalone or as a plug-in of the LArSoft software suite~\cite{Snider:2017wjd} used by many LArTPC experiments.  This simulation module is the focus of current study. 

\section{Initial Porting with CUDA}
\label{sec-cuda}

\begin{figure}[thb]
  \centering
  \includegraphics[width=1.0\figwidth]{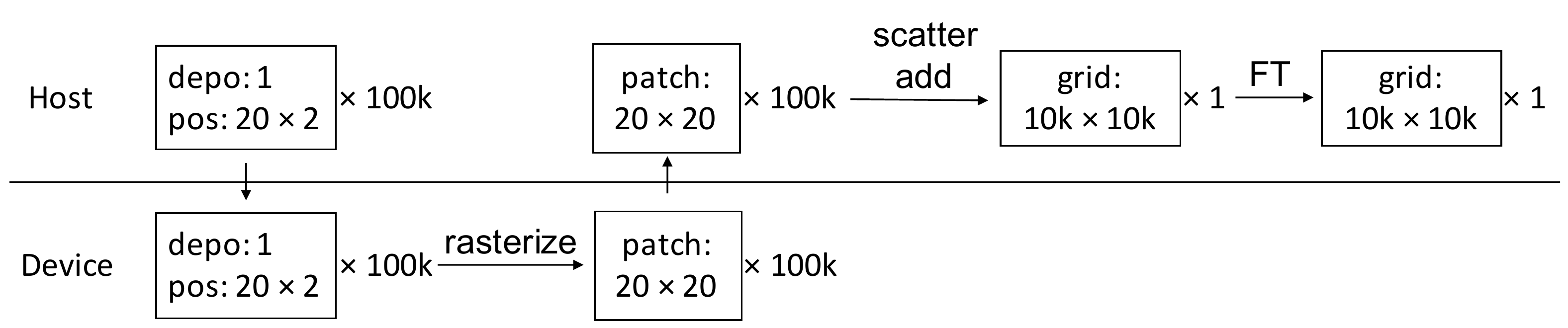}
  \caption{Data flow for current CUDA porting and first round of Kokkos porting.}
  \label{fig:data-flow-current}
\end{figure}

Before adopting Kokkos in \wc, we evaluated the potential of using NVIDIA GPUs to accelerate the signal simulations using CUDA. Performance profiling of the CPU code revealed that the most time-consuming part is in the rasterization step, which is what we ported to CUDA first. The data flow for this initial porting is shown in Figure~\ref{fig:data-flow-current}, and is largely based on the original CPU code. In this example, we compute 100,000 (100k) energy depositions (depos) with a patch size of $20\times20$ each. The energy depositions  are transferred to the device from the host one at a time, rasterized on the device and then transferred back to the host. The concurrency on the device in this scenario is very low ($20\times20$) and only 1 GPU thread block is used. This is only a partial porting, and we do not expect the performance to be good at this stage.  There are three reasons why the initial strategy does not yield a good performance. First, the data need to be transferred back and forth for the rasterization of each patch, incurring significant data transfer overhead. Second, the size of the patch to be computed on the GPU is very small, only $20\times20$, resulting in the significant under-utilization of the GPU. Third, there are still two  parts of the simulation, ``scatter add" and ``FT", that are computed serially, limiting the overall performance gain we can achieve. We will see the effects of these limitations in the performance benchmarking results presented in Section~\ref{subsec-kokkos-results}.

\begin{figure}[thb]
  \centering
  \includegraphics[width=1.0\figwidth]{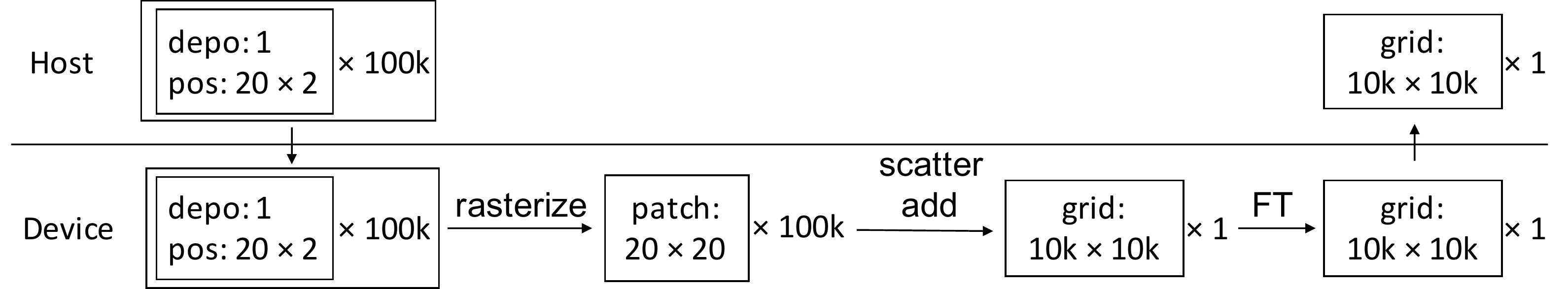}
  \caption{Planned data flow for the final Kokkos porting.}
  \label{fig:data-flow-plan}
\end{figure}

This initial porting is done to minimize the code changes needed, and serves as a way for us to familiarize ourselves with the CUDA porting process. And we have a plan to further improve the performance of the code by following the strategy in Figure~\ref{fig:data-flow-plan}, where all three steps in Eq.~\ref{eq:conv-freq} will be computed on GPUs. This will also allow us to only do the data transfer once at the beginning of the simulation and once at the end, which will help reduce the data transfer overhead and improve the amount of computation performed on the GPU. But this strategy requires more significant code restructuring, which is in progress.  

%
\section{Portable Solution with Kokkos}
\label{sec-kokkos}

\subsection{Introduction to Kokkos}
\label{subsec-kokkos-intro}
Kokkos~\cite{kokkos} is a C++ library that allows developers to write code for different computing architectures and parallelism paradigms using a single API. In addition to supporting standard parallel algorithms, including a tasking model, Kokkos also supports diverse node architectures and memory models by allowing users to define their own execution and memory spaces.  

The Kokkos abstraction layer maps C++ source code to the specific instructions required for the Kokkos \texttt{backends} enabled during build time.  When compiling the source code, the binary code for up to three backends may be generated for a given C++ translation unit:

\begin{itemize}
\item \texttt{Serial} backend, which executes single-threaded on a host device.
\item \texttt{Host-parallel} backend, which executes multithreaded on the host device.
\item \texttt{Device-parallel} backend, which executes on an external device (e.g. a GPU).
\end{itemize}

\noindent For the host-parallel backends, Kokkos currently supports multi- and many-core CPUs through OpenMP~\cite{omp} or POSIX threads (pthreads).  The device-parallel backends include CUDA for NVIDIA GPUS, and HIP for AMD GPUs. In the latest version of Kokkos, experimental support for OpenMP target offloading and SyCL has also been added. 
 
\subsection{Software infrastructure to support Kokkos porting}

In order to get Kokkos to build with \wct, there are several changes to software infrastructure we had to make to enable and facilitate Kokkos porting and testing. These include a Kokkos-enabled container environment and changes to the \wct framework, which will be discussed in more detail below.


\subsubsection{Containerized development and production}
\label{subsec-kokkos-container}

Although input data can be presented to \wct in its standalone form via JSON serialization, more realistic data can be provided through the \larsoft library~\cite{Snider:2017wjd}.  The \larsoft packages and their software dependencies require consistent versions and compatibly-built binary libraries.  As the explorations described here must be performed on various platforms, a development solution portable to multiple platforms was necessary.

To achieve this, the \larsoft packages (and their software dependencies), \wct, Kokkos library, and relevant Kokkos backends were included in a CentOS-7 based Docker image.  Software products and their versions can be found in Table~\ref{tab:software}.  The Docker images can be readily converted to Singularity and Shifter images, the latter being required by NERSC's Cori platform.  Upon starting a container of the image, an entrypoint script initializes the environment, making available all \larsoft, \wc, and Kokkos libraries needed for compiling and running (within the container) the code described in this paper.

\begin{table}[ht]
\centering
\caption{Software packages and their corresponding versions used for this study.} 
\label{tab:software}
\begin{tabular}{ll}
\hline
Software package            & Version \\ \hline
GCC       & 8.2 \\
Kokkos    & 3.3.00 \\
CUDA toolkit & 11.0.2 \\
Eigen & 3.3.9 (with patches) \\ \hline
\end{tabular}
\end{table}
\subsubsection{Framework development}
\label{subsec-kokkos-framework}

To be able to develop relative independently with the \wct production code base, we created a standalone package \texttt{wire-cell-gen-kokkos}~\cite{wct-gen-kokkos}.
It produces a shared library which could be used by \wct as a plugin.
We can use two building systems based on \texttt{waf}~\cite{waf} and \texttt{cmake}~\cite{cmake}.
To accommodate Kokkos code without affecting the existing C++ and CUDA code, we use a customised file extension ``.kokkos'' for the source code using Kokkos so that the compiling system can easily identify them and treat them differently if needed.
To acquire and release resources needed by Kokkos, \texttt{Kokkos::initialize()} and \texttt{Kokkos::finalize()} need to be called before and after the actual Kokkos code executes.
A \texttt{KokkosEnv} object inherited from \texttt{WireCell::ITerminal} was created to handle this.  The \texttt{Kokkos::initialize()} is called in its constructor and the \texttt{Kokkos::finalize()} is called when the \texttt{WireCell::ITerminal::finalize()} is called for a stack of \texttt{WireCell::ITerminal} objects before the program exits. We note that this may be the unique feature and requirement of \wct, given its dataflow programming model. 
\subsection{Kokkos porting}

\subsubsection{Charge rasterisation kernel}
\label{subsec-kokkos-rast}

As mentioned in Section \ref{sec-cuda}, the current CUDA porting focuses on the rasterization of individual energy depositions as shown in Figure~\ref{fig:data-flow-current}.
For the first round of Kokkos porting, we decided to follow Figure~\ref{fig:data-flow-current} and only port the rasterization part which already has a CUDA version.
The advantage of this strategy is that we can have a direct comparison between Kokkos and CUDA in terms of performance and development effort. We will eventually finish porting the whole simulation as shown in Figure~\ref{fig:data-flow-plan}. 

We have discovered that there are some missing components in Kokkos are needed for \wct. For example, \wct uses random numbers with a normal distribution or binomial distribution. There does not seem to be a Kokkos function that does this. We resorted to using the Box-Muller transform \cite{box1958note} to generate normal-distribution random numbers from uniformly distributed ones. Similar to the CUDA implementation,  we implemented a random number pool to allow  multiple threads to access the random numbers concurrently.


\subsubsection{Preliminary benchmark results}
\label{subsec-kokkos-results}

We tested three LArTPC simulation implementations (original CPU, CUDA and Kokkos) on our workstation with one 24-core AMD Ryzen Threadripper 3960X CPU and one NVIDIA V100 GPU. The tests were done in the container environment as described in Section~\ref{subsec-kokkos-container}, with the software information in Table~\ref{tab:software}.
The original serial CPU implementation and CUDA implementation described in Section~\ref{sec-cuda} will be referred to as ``ref-CPU" and ``ref-CUDA" later.
For the Kokkos implementation, we tested two backends, the OpenMP backend and CUDA backend, referred to as ``Kokkos-OMP" and ``Kokkos-CUDA" later.
The input for the tests are energy depositions generated from simulated cosmic rays interacting with liquid argon (LAr).
We used \texttt{CORSIKA}~\cite{Engel:2018akg} as the cosmic ray generator and \texttt{Geant4}~\cite{AGOSTINELLI2003250,1610988,ALLISON2016186} to simulate the particle and LAr interactions.
The software stack used to to generate the input files is \texttt{LArSoft}~\cite{Snider:2017wjd}. 

Table~\ref{tab:result-ref-cpu-cuda} and Table~\ref{tab:result-kokkos} summarize the preliminary timing results for the rasterization part that has been ported to CUDA and Kokkos following strategy in Figure~\ref{fig:data-flow-current}.
The second column from the left listed the total rasterization time, while the third and forth listed timing for two major steps of the rasterization. We ran each test 5 times and averaged the timing results, but the fluctuations between runs were fairly small. 

From the top two rows of Table~\ref{tab:result-ref-cpu-cuda} we can see the total rasterization time is reduced by about factor of 3 using CUDA.
However, we can see most of the speed-up comes from the fourth column, the fluctuation calculation.
This step contains running a random number generator (RNG) in the ref-CPU implementation.
Current realization of the RNG in ref-CPU is \texttt{std::binomial\_distribution}.
In ref-CUDA, the RNG is factored out from the fluctuation calculation, and instead a pre-calculated random number pool is used.
When we temporarily remove the RNG from ref-CPU, we get results in the third row, referred to as ref-CPU-noRNG. 
Comparing ref-CUDA and ref-CPU-noRNG, we can see there is no speedup, but rather a slowdown, with GPU in this round of implementation.
A detailed analysis shows that there are three reasons for the poor performance:
\begin{enumerate}
\item Data are transferred between host and device in many blocks of a few kilobytes, unnecessarily increasing the data transfer cost; 
\item The number of units of workload that needs to be parallelized (concurrency) is too small (< 1000);
\item The amount of work per unit is too small compared to the dispatch overhead.
\end{enumerate}
Based on this result and analysis, we proposed a strategy that could improve the performance as described in Figure~\ref{fig:data-flow-plan}.
In this strategy, the data blocks copying to device will be batched to improve the transfer efficiency.
This also increases concurrency significantly.
The relative dispatch overhead could also be reduced.
Most importantly, in this strategy, the data could stay on the device for the next two major steps (scatter-add and FT).
Data copying time would be further reduced.

Table~\ref{tab:result-kokkos} shows timing results for the current Kokkos implementation, also following Figure~\ref{fig:data-flow-current}. Note that the RNG has been pre-calculated, similar to the CUDA implementation, and is thus not included in the timing for the Kokkos implementation. 
For Kokkos-OMP, with an increased number of maximum threads, the computing time actually gets longer.
This is a sign that the dispatch overhead outweighs the parallelization benefit.
We think this could also be improved by adopting the Figure~\ref{fig:data-flow-plan} strategy.
Comparing Kokkos-CUDA and ref-CUDA, we can see that Kokkos-CUDA is about two times slower than ref-CUDA.
Analyses using the NVIDIA Nsight Systems~\cite{nsys} show that the causes are: 
1, Kokkos \texttt{parellel\_reduce()} kernels are almost 3 times slower than CUDA reduction kernels;
2, in between kernel and API calls, Kokkos has extra \texttt{CudaDeviceSynchronization} and \texttt{CudaStreamSynchronization}. For a workflow with many small calculation kernels and data transfers, those extra steps have significant contributions to the run time.

\begin{table}[ht]
\centering
\caption{Preliminary timing results  for the original serial CPU implementation and CUDA implementation described in Section~\ref{sec-cuda}. Unit is second (s). For ref-CUDA, the timing includes data transferring time between device and host. ``h->d" means data transfer from host to device; ``d->h" the other way around.}
\label{tab:result-ref-cpu-cuda}
\begin{tabular}{cccc}
\hline
Description                & Rasterization total [s]    & 2D sampling [s]        & Fluctuation [s] \\\hline
ref-CPU                    & 3.57                    & 0.07                & 3.42 (incl. RNG)\\
ref-CUDA                   & 1.22                    & 0.21 (incl. h->d)      & 0.79 (no RNG, incl. d->h)\\
ref-CPU-noRNG          & 0.18                    & 0.07                & 0.03 (no RNG)\\\hline
\end{tabular}
\end{table}

\begin{table}[ht]
\centering
\caption{Preliminary timing results  of the first round Kokkos porting using strategy in Fig.~\ref{fig:data-flow-current}. Unit is second (s). Note that the Kokkos implementation uses a pre-calculated random number pool, so the timing should be compared to ref-CUDA and ref-CPU-noRNG in Table~\ref{tab:result-ref-cpu-cuda}.}
\label{tab:result-kokkos}
\begin{tabular}{cccc}
\hline
Description            & Rasterization total [s]    & 2D sampling [s]        & Fluctuation [s] \\\hline
Kokkos-OMP  1 thread    & 0.29                    & 0.11                & 0.10\\
Kokkos-OMP  2 thread    & 0.49                    & 0.17                & 0.23\\
Kokkos-OMP  4 thread    & 0.55                    & 0.19                & 0.28\\
Kokkos-OMP  8 thread    & 0.66                    & 0.24                & 0.34\\
Kokkos-CUDA             & 2.31                    & 0.94                & 1.28\\\hline
\end{tabular}
\end{table}

\section{Future Plans}
\label{sec-kokkos-plan}

As discussed in Section~\ref{sec-cuda}, three major algorithms need to be ported to Kokkos: the rasterization, scatter adding and Fourier Transform.
The rasterization has a CUDA implementation and porting to Kokkos is relatively easy, as discussed in Section~\ref{subsec-kokkos-rast}. However, the performance is not ideal as we discussed in Section~\ref{subsec-kokkos-results}. We have identified the causes for the poor performance and will improve our Kokkos implementation according to the discussions in Section~\ref{subsec-kokkos-results}. 

\begin{figure}[thb]
  \centering
  \includegraphics[width=0.6\figwidth]{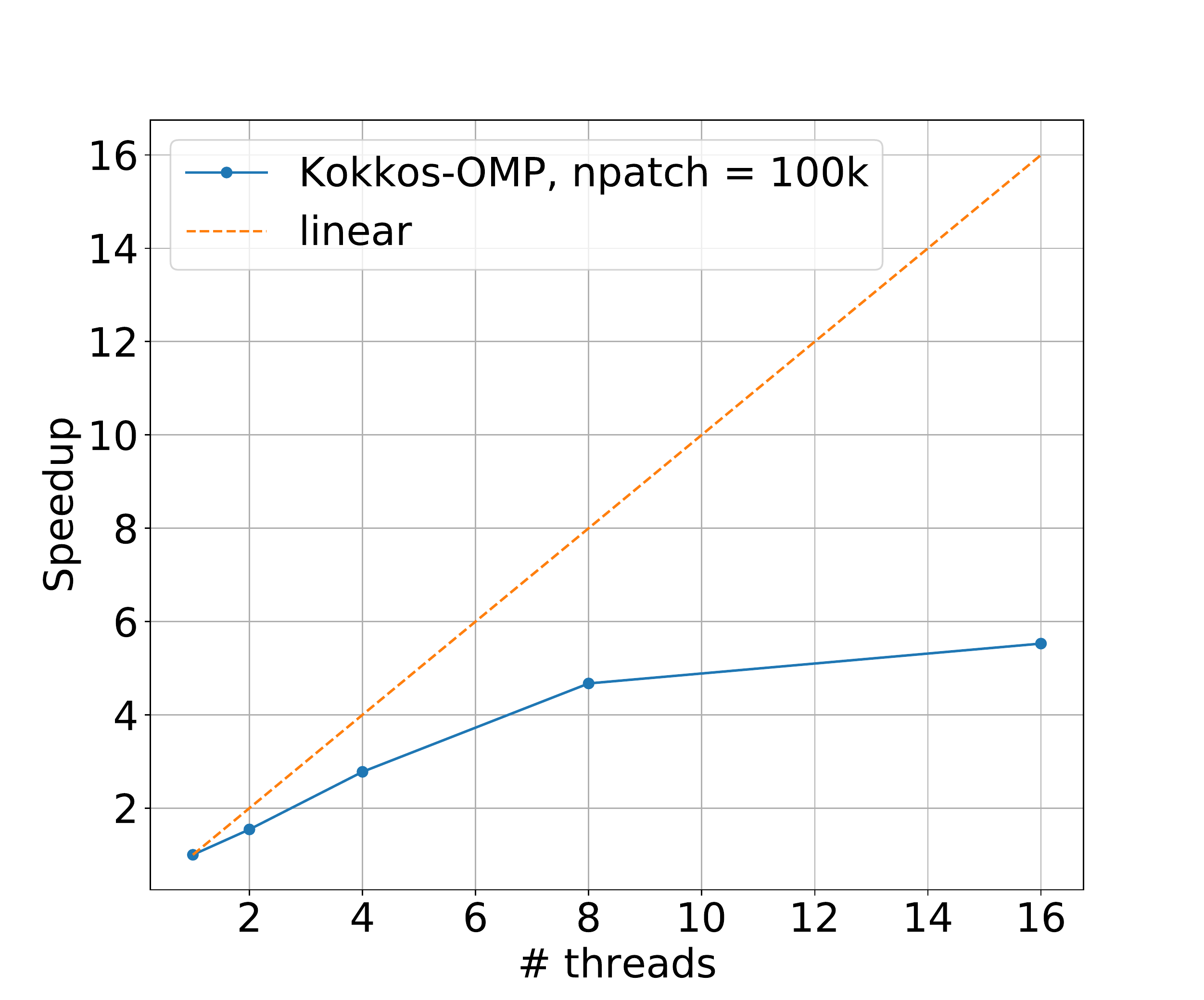}
  \caption{Scalability test for the scatter adding task using \texttt{Kokkos::atomic\_add}. The test was performed on a machine with an Intel i9-9900K CPU (8 cores).
  }
  \label{fig:scatter-add-prof}
\end{figure}

For the scatter-adding component, we will use \texttt{Kokkos::atomic\_add}. We have implemented a unit test for \texttt{Kokkos::atomic\_add} in \emph{wire-cell-gen-kokkos} to test the correctness and performance of our implementation.
An initial scalability test using Kokkos with OpenMP backend is shown in Figure~\ref{fig:scatter-add-prof}. The speedup is relative to the serial CPU reduction. Since the test machine only has 8 CPU cores, the flattening of the curve simply reflects that the compute capacity has been exhausted after 8 threads. 

For the Fourier Transform we are currently using \textit{Eigen}~\cite{eigen} with \textit{fftw}~\cite{fftw} Fast Fourier Transform (FFT) backend in the CPU code.
Right now, Kokkos does not have a native FFT implementation or common interface for various optimized vendor FFT libraries. Until Kokkos adds support for FFT, we will implement our own wrapper APIs over the vendor FFT libraries for different backends, similar to the approach taken by the Synergia~\cite{synergia} team. 
This work is ongoing.

\section{Summary}
\label{sec-summary}


We have explored the feasibility of using Kokkos to implement a portable acceleration solution for LArTPC simulation in \wct. 
We have ported the original serial CPU implementation to CUDA and then to Kokkos following a simple and localized strategy without significantly refactoring the original CPU code.
During this process, we have learned that factoring out a random number calculation from the main loop could significantly reduce the computing time.
However, this localized strategy does not seem enough to efficiently use the GPU accelerators due to very low concurrency.
As a result, we have proposed an alternative porting strategy which needs more code refactoring but should be able to better utilize accelerators.
The Kokkos implementation does run on the two different backends we tested (OpenMP and CUDA) without any change of the source code.
Its performance with CUDA backend degrades non-negligibly from the raw CUDA implementation in our test.
Given it is our initial Kokkos implementation, it is very likely that we can further optimize it in our future work.

\section*{Acknowledgments}
This work was supported by the U.S. Department of Energy, Office of
Science, Office of High Energy Physics, High Energy Physics Center for
Computational Excellence (HEP-CCE) at Brookhaven National Laboratory and
Fermi National Accelerator Laboratory under B\&R KA2401045.  This research used resources of the National Energy Research Scientific Computing Center (NERSC), a U.S. Department of Energy Office of Science User Facility located at Lawrence Berkeley National Laboratory, operated under Contract No. DE-AC02-05CH11231. We gratefully acknowledge the use of the computing resources at the Scientific Data and Computing Center of Brookhaven National Laboratory, as well as the support of the Wire-Cell team of the Electronic Detector Group in the Brookhaven National Laboratory Physics department, both of which are supported by the U.S. Department of Energy under Contract No. DE-SC0012704.

\bibliography{main}

\end{document}